# Absorption of Carbon Dioxide in Kerogen Nanopores: A Mechanism Study using the Molecular Dynamics–Monte Carlo Method

Jie Liu[1], Tao Zhang[1*], Shuyu Sun[1*]

1 Physics Science and Engineering, King Abdullah University of Science and Technology, Thuwal, Saudi Arabia, 23955-6900

[*]Corresponding Author: tao.zhang.1@kaust.edu.sa, shuyu.sun@kaust.edu.sa

**ABSTRACT**

Carbon capture and storage (CCS) technology has been applied successfully in recent decades to reduce carbon emissions and alleviate global warming. In this regard, shale reservoirs present tremendous potential for carbon dioxide ($CO_2$) sequestration as they have a large number of nanopores. Molecular dynamics (MD) and MD–Monte Carlo (MDMC) methods were employed in this work to study the absorption behavior of $CO_2$ in shale organic porous media. The MDMC method is used to analyze the spatial states of $CO_2$, and the results are in good agreement with MD's results, and it also performs well in the acceleration compared to the classical MD. With regard to the kerogen matrix, its properties, such as the pore size distribution (PSD), pore volume, and surface area, are determined to describe its different compression states and the effects of $CO_2$ absorption on it. The potential energy distribution and potential of mean force are analyzed to verify the spatial distribution of $CO_2$ molecules. The heterogeneity of the pore structure resulted in heterogeneous distributions of $CO_2$ molecules in kerogen porous media. Moreover, strong compression of the matrix reduces the number of large pores, and the PSD is mainly between 0 and 15 Å. Despite the high interaction force of the kerogen matrix, the high-potential-energy region induced by the kerogen skeleton also contributes to the formation of low-energy regions that encourage the entry of $CO_2$. An increase in temperature facilitates the absorption process, allowing $CO_2$ molecules to enter the isolated pores with stronger thermal motion, thereby increasing the storage capacity for $CO_2$. However, the development of geothermal energy may not be suitable for $CO_2$ sequestration.

# 1 Introduction

In the last decades, carbon capture and storage (CCS) has been considered an effective technique for alleviating global warming caused by excess carbon emissions [1, 2]. The most common methods for realizing successful CCS are geological sequestration, ocean storage, and natural capture and storage by plants [3]. Among these, geological sequestration has been widely used in many countries because of its excellent performance in storing carbon dioxide ($CO_2$) [4]. Although, petroleum and natural gas molecules are known to exist in rock porous media as free and adsorbed phases, studies on the absorption behavior of $CO_2$ molecules in shale organic nanopores have not been conducted so far, preventing a comprehensive understanding of $CO_2$ sequestration.

Different types of reservoirs have been utilized for CCS projects, including shale, carbonate, and sandstone reservoirs [5-7]. Among these, shale rocks have extremely low porosity and numerous nanopores, yielding a large surface area for adsorption [8]. Thus, efforts have been made to compare the adsorption capacities of $CO_2$ and shale gas [9, 10], and their results reveal that $CO_2$ has greater affinity for the shale matrix than the shale gas. Furthermore, shale's extremely tight rock structure act as the gap rock to prevent $CO_2$ leakage, but organic matter can migrate into nanopores in a long-term stratigraphic evolution [11]. Consequently, the effect of $CO_2$ on the changes in organic nanostructures of shale is crucial for the sealing layer. Not only does the shale reservoirs exhibit great potential for storing and sealing $CO_2$, but $CO_2$ has been widely used to enhance oil recovery [12, 13]. Therefore, $CO_2$ and shale interaction research is crucial for various engineering applications.

Organic matter is commonly found in shale reservoirs because of its self-generating and self-storing properties [14]. Oil and natural gas are derived from kerogen, which represents the majority of organic matter [15, 16]. In this regard, Bousige et al. [17] built realistic kerogen molecular models using a hybrid experimental–simulation method and adopted the $CO_2$ adsorption experiment to probe micropores in the kerogen matrix, where the pore size distribution (PSD) was mostly below 10 Å. Meanwhile, Ungerer et al. [18] classified various kerogen types on the basis of maturities and proposed corresponding molecular models, and the density was set as 1.0 g·cm$^{-3}$ for the type I kerogen matrix. Moreover, Zhang et al. [19] conducted simulations on oil shale with a kerogen density of 1.0 g·cm$^{-3}$. By performing simulations with a kerogen density of 0.95 g·cm$^{-3}$, Facelli et al. [20] demonstrated good agreement between the results of simulation and experimental measurements using actual samples. Therefore, the appropriate modeling of the kerogen matrix must be carefully considered to obtain accurate results.

Although $CO_2$ has already been utilized in the development of shale energy for decades, $CO_2$-related problems in nanopores remain a major challenge that is often encountered in laboratories. Molecular dynamics (MD) simulation, a particle method based on the Lagrange coordinate [21, 22], is capable of tackling the interaction behaviors and mechanism explanation at the nanoscale, and this method has been applied to some kerogen-related and $CO_2$-related problems successfully in recent years [23-26]. In this regard, Perez et al. [27] studied the adsorption of multicomponent fluids within the kerogen matrix via molecular simulations. Their findings indicated that $CO_2$ molecules are preferentially adsorbed onto the surface of kerogen pores and diffused into the kerogen inner matrix. Meanwhile, Psarras et al. [28] investigated the $CO_2$ storage capacity in kerogen using molecular simulations by measuring the surface properties of nanopores with diameters below 15 Å. $CO_2$ was shown to have high-priority adsorption position in rigid kerogen matrixes at high temperatures [23, 29]. However, thermal motion may have a significant impact on the absorption of $CO_2$, and absorbed $CO_2$ molecules may also affect the structural properties of the matrix. As a consequence, it is imperative that the flexibility of the kerogen matrix and $CO_2$'s effects must be considered for more reasonable simulations.

MD simulations have the capability of generating accurate results. However, rare events, such as the presence of $CO_2$ within some unconnected nanopores, are always beyond the MD time scale. Every spatial state of $CO_2$ in the kerogen matrix can be reached by consuming a large amount of computational resources. Thus, the MD–Monte Carlo (MDMC) method, described in our previous study [30], was implemented to speed up the above molecular simulations and to reach rare states at high frequency. MD results from long-term calculations serve as the ground truth for calculating the probability of each state; this enables the prediction of transition between different states. The MDMC algorithm can implement coarsening operations to accelerate the calculation at the spatial and temporal scales. Consequently, the MDMC method sacrifices a small amount of spatial state distribution details in favor of computational speed.

In this study, MD and MDMC methods are adopted to investigate $CO_2$ sequestration in kerogen nanopores at the nanoscale. Spatial distributions in different dimensions can be simulated and matched using these two methods. Furthermore, spatial and temporal coarsening operations are employed to accelerate the computational speed of the MDMC algorithm. Additionally, the PSD, pore volume, and surface area of the kerogen matrix are investigated at different compression conditions. Moreover, the absorption of $CO_2$ and its effects on the properties of kerogen porous structure are examined. The energy mechanism governing the absorption behavior of $CO_2$ in the kerogen matrix at various temperatures is also explained.

## 2 Methodology

### 2.1 Molecular models

As depicted in Fig. 1, $CO_2$ molecules, dummy particles, kerogen monomers, and graphene layers were modeled separately to build the simulation system. Twenty-seven kerogen monomers were packed in the center region of the system, as displayed in Fig. 2, and $CO_2$ molecules filled two sides, where the $CO_2$ and kerogen matrix were separated by four graphene plates. The types of kerogen were mainly types I and II in China [31, 32]. Herein, the type I-A kerogen model ($C_{251}H_{385}O_{13}N_7S_3$) was adopted as it had good stability during the simulation. The dummy particle was utilized to promote the formation of large pores in the kerogen matrix [33], and the distance parameter was set at 1.5 nm in the force field. Graphene plates played a role of compressing the kerogen matrix and driving the $CO_2$. In addition, all graphene layers were kept as rigid bodies to obtain a more stable computational performance during the simulation, and the size of the graphene layer was kept the same as the box size in *X* and *Y* directions. Specific fractions of different components are listed in Table 1. The size of the simulation box was $5.2 \times 5.2 \times 200$ Å$^3$.

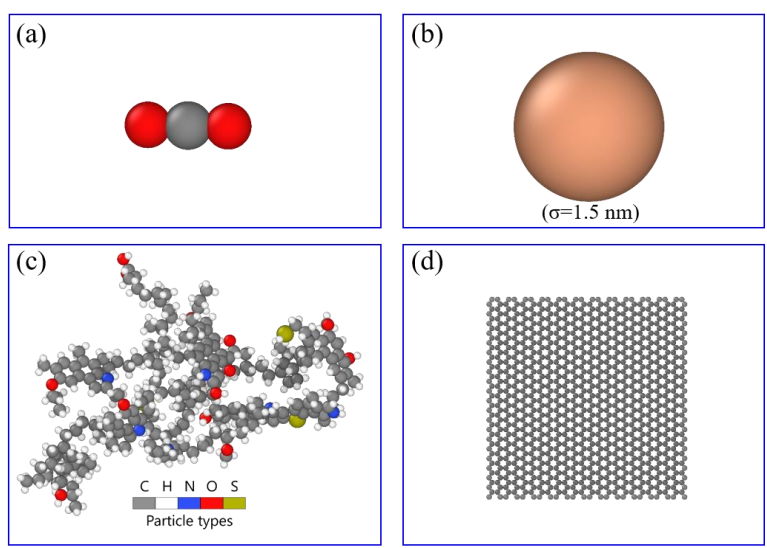

Fig. 1 Molecular models: (a) $CO_2$, (b) dummy particle, (c) type I-A kerogen monomer ($C_{251}H_{385}O_{13}N_7S_3$), and (d) graphene plate.

Table 1 Molecular fractions of different components.

| Species | Atomic number in one molecule | Number of molecules | Total atoms of each component |
|---|---|---|---|
| $CO_2$ | 3 | 582 | 1746 |
| Graphene | 1008 (in one layer) | 4 (layers) | 4032 |
| Kerogen monomer | 659 | 27 | 17793 |
| Dummy particle | 1 | 9 | 9 |

The molecular interaction parameters were described by the polymer consistent force field [24, 34], as summarized in Table 2, and parameters between different atomic types were determined using Waldman–Hagler combining rules [35]. Moreover, the energy parameters of the graphene layers were modified to avoid too strong interaction forces between the graphene layers and kerogen molecules. With the use of force field parameters and the Lennard–Jones equation [36], Van der Waals interactions were calculated with respect to different atoms. The electrostatic force was addressed by means of the Ewald method [37]. Periodic boundary conditions were applied in all directions, and the cutoff distance was set to 1.2 nm in this study.

Table 2 Molecular parameters of different atoms.

| Species | Atom type | Mass [g·mol$^{-1}$] | Charge [e] | $\varepsilon$ [kJ·mol$^{-1}$] | $\sigma$ [nm] |
|---|---|---|---|---|---|
| $CO_2$ | c=2 | 12.011 | 0.24 | 0.064 | 0.401 |
| | o= | 15.9994 | −0.12 | 0.06 | 0.3535 |
| Graphene | cp | 12.011 | 0.0 | 0.02 | 0.401 |
| dummy particle | dummy | 12.011 | 0.0 | 0.002 | 1.5 |
| Kerogen matrix (type I-A) | c2 | 12.011 | −0.106 | 0.054 | 0.401 |
| | hc | 1.00794 | 0.053 | 0.02 | 0.2995 |
| | c5 | 12.011 | −0.05 | 0.064 | 0.401 |
| | nh | 14.00674 | −0.2925 | 0.134 | 0.407 |
| | oc | 15.9994 | −0.1595 | 0.24 | 0.3535 |
| | na | 14.00674 | −0.2481 | 0.065 | 0.407 |
| | c1 | 12.011 | −0.053 | 0.054 | 0.401 |
| | c3 | 12.011 | −0.159 | 0.054 | 0.401 |
| | cs | 12.011 | −0.013 | 0.064 | 0.401 |
| | sp | 32.064 | 0.026 | 0.071 | 0.4027 |
| | np | 14.00674 | −0.481 | 0.041 | 0.357 |
| | c | 12.011 | 0.0 | 0.054 | 0.401 |
| | c_0 | 12.011 | 0.3964 | 0.12 | 0.3308 |
| | c_1 | 12.011 | 0.702 | 0.12 | 0.381 |
| | o_1 | 15.9994 | −0.531 | 0.267 | 0.33 |
| | sh | 32.064 | −0.2042 | 0.071 | 0.4027 |
| | hs | 1.00794 | 0.1392 | 0.02 | 0.2995 |
| | o_2 | 15.9994 | −0.594 | 0.24 | 0.342 |
| | ho2 | 1.00794 | 0.423 | 0.013 | 0.111 |
| | hn | 1.00794 | 0.3925 | 0.013 | 0.1098 |
| | oh | 15.9994 | −0.5571 | 0.24 | 0.3535 |

| | ho | 1.00794 | 0.4241 | 0.013 | 0.1098 |
| | sc | 32.064 | −0.13 | 0.071 | 0.4027 |

## 2.2 Traditional MD simulations

MD simulations were addressed using a large-scale atomic/molecular massively parallel simulator package [38], and visualization was performed by taking advantage of the OVITO software [39]. The properties of kerogen porous media, such as PSD, surface area, and pore volume, were figured out by means of the Zeo++ package [40-42]. The MD simulations in this work were separated into two parts: the compression of the kerogen matrix and the adsorption of $CO_2$ to the kerogen nanopores. Furthermore, the kerogen matrix was thermally controlled to keep its flexibility for obtaining a highly realistic physical porous media. Although the frozen kerogen matrix could lead to a more stable simulation, it would also induce nonphysical resistance to the injected fluid; this shortcoming would be more pronounced for the absorption problem. The specific information regarding the compression of the kerogen matrix is shown in Table 3. The Nosé–Hoover thermostat method was adopted to control the temperature [43, 44].

Table 3 Processes of kerogen matrix compression.

| State of compression | Distance of compression [Å] | Time step [fs] | Simulation time [ps] | Starting temperature [K] | End temperature [K] |
|---|---|---|---|---|---|
| Relaxation | 0 | 0.1 | 20 | 300 | 330 |
| State 1 | 6 | 0.2 | 1.5 | 330 | 330 |
| Relaxation | 0 | 0.5 | 500 | 330 | 330 |
| State 2 | 6 | 0.2 | 1.5 | 330 | 330 |
| Relaxation | 0 | 0.5 | 500 | 330 | 330 |
| State 3 | 6 | 0.2 | 1.5 | 330 | 330 |
| Relaxation | 0 | 0.5 | 500 | 330 | 330 |
| State 4 | 6 | 0.2 | 1.5 | 330 | 330 |
| Relaxation | 0 | 0.5 | 500 | 330 | 330 |
| State 5 | 6 | 0.2 | 1.5 | 330 | 330 |
| Relaxation | 0 | 0.5 | 500 | 330 | 330 |
| State 6 | 6 | 0.2 | 1.5 | 330 | 330 |
| Relaxation | 0 | 0.5 | 500 | 330 | 330 |

*In the case of target temperature = 330 K.

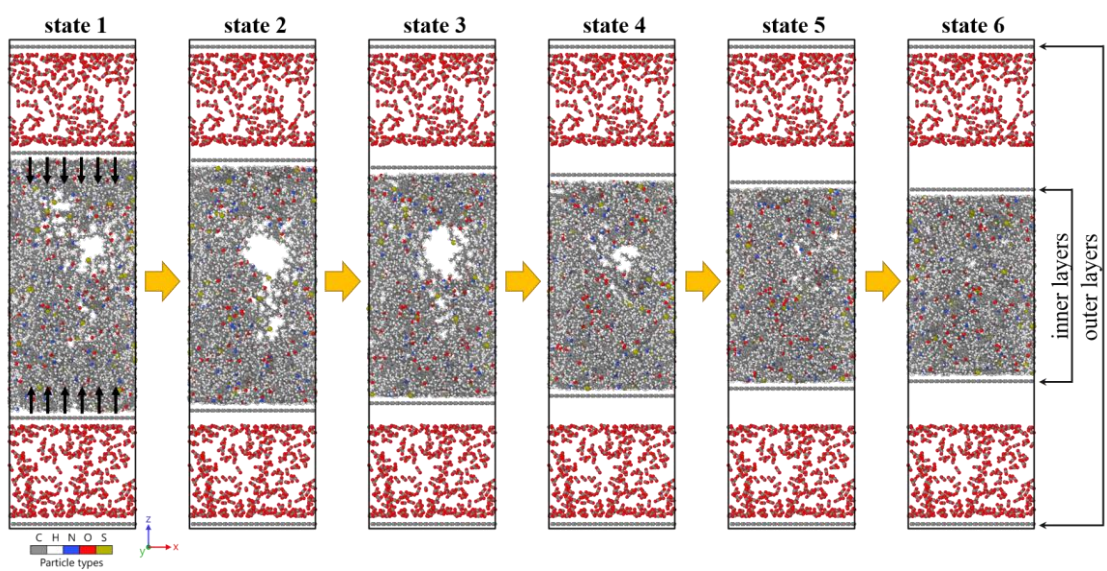

Fig. 2 Different states of compression of the kerogen matrix. Black arrows represent the compression direction and four graphene layers are classified into two inner layers and two outer layers.

First, the kerogen matrix was compressed by two inner graphene layers, as visualized in Fig. 2. The $CO_2$ component was not thermally controlled during these processes to avoid its influence on the kerogen matrix. These processes were performed to mimic the kerogen matrix in different reservoir conditions, which are usually embodied in the depths of reservoirs [45]. In shale reservoirs, rocks are extremely tight and organic matter is usually highly compressed [46]. $CO_2$ absorption simulation was started once the kerogen matrix reached its target state. The specific absorption process of $CO_2$ is shown in Table 4. The $CO_2$ absorption was achieved by pushing two outer layers (Fig. 3) until the graphene layers were coincident. Additionally, the neighbor contribution between the inner and outer layers was neglected, which made it possible for the atom overlapping. The interaction force was also canceled between $CO_2$ molecules and inner layers; hence, these molecules were able to transport across the inner graphene layers. The relaxation simulation was exerted to obtain the optimized $CO_2$ spatial distribution in the kerogen porous media. Thereafter, the result could be collected during a long-time simulation.

Table 4 Simulation processes of $CO_2$ injection and absorption.

| State of compression | Time step [fs] | Simulation time [ps] | Starting temperature [K] | End temperature [K] |
|---|---|---|---|---|
| Target state relaxation | 0.5 | 500 | 330 | 330 |
| $CO_2$ injection | 0.2 | 277 | 330 | 330 |
| Relaxation | 0.5 | 1000 | 330 | 330 |
| Data collection | 0.5 | 1500 | 330 | 330 |

*In the case of target temperature = 330 K.

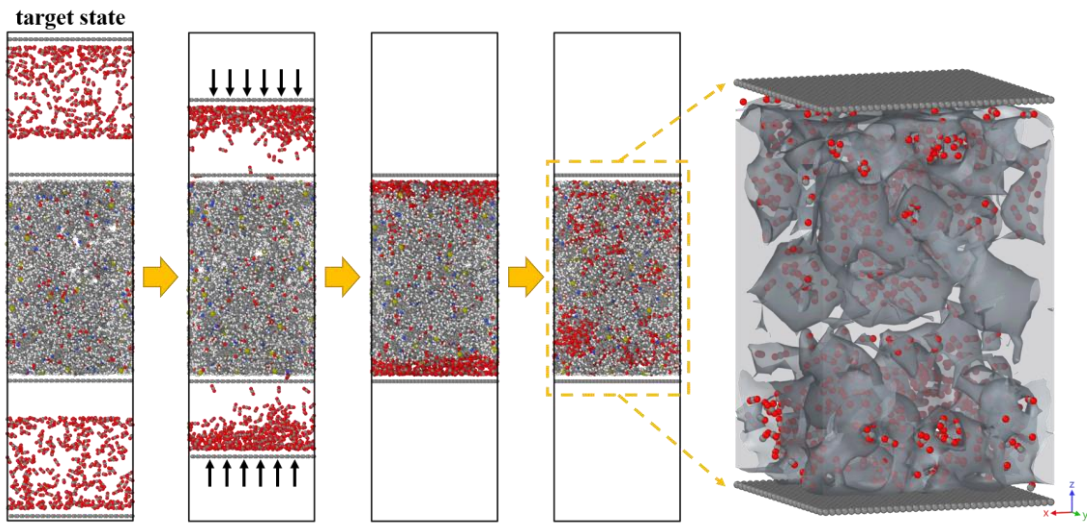

Fig. 3 $CO_2$ absorption process in the kerogen porous media, where the transparency gray region represents the pore region.

## 2.3 MDMC simulations

The MD simulation was conducted by adopting specific potential functions to predict the next position of each atom, and the position along the time series formed the atomic trajectory. If each spatial point of the atom was regarded as a state, the trajectory could be considered as the state distribution in space. The overlapped parts of different trajectories presented a spatial state with high frequency, which meant that the trajectory could be converted to the spatial probability distribution. Consequently, the MC process could be presented using the probability distribution yield from MD simulations [30]. In doing so, given the high accuracy of the MD simulations, the result can be viewed as the ground truth of the MDMC algorithm.

The coarsening operation in the MDMC algorithm was conducted to accelerate the computational speed at the spatial and temporal scales within an acceptable margin of error. By discretizing the computational domain with a larger mesh size, the number of spatial states was reduced, which led to a smaller size of the probability transition matrix, and spatial coarsening could be realized. Meanwhile, temporal coarsening was addressed with a larger time step. In this way, the computation could be accelerated by the corresponding time multiplier. Fig. 4 presents the schematic workflow of the MDMC algorithm. The size of the probability transition matrix was $N \times N$, where $N$ was the total state number of corresponding problems. Specific details of the MDMC algorithm were introduced and validated in our previous work [30].

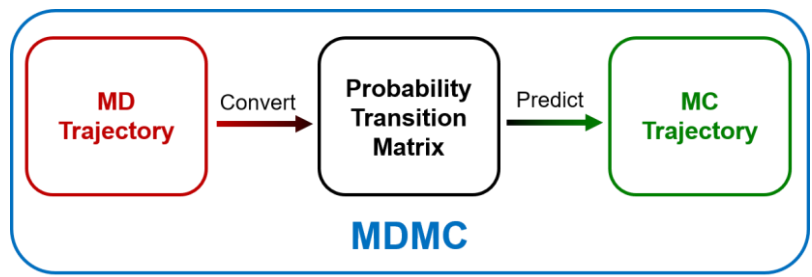

(a)

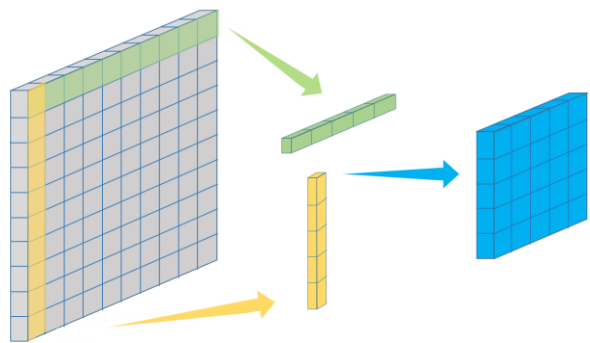

(b)

Fig. 4 Workflow of (a) the MDMC method and (b) coarsening operation on the probability transition matrix.

## 3 Results and discussions

### 3.1 Spatial distribution characteristics of $CO_2$ from the MD and MDMC methods

Spatial distribution is the most basic and intuitive form of expression that can be used to represent MD results and can be used as a basis for calculating other properties. Fig. 5 displays the density profiles of kerogen and $CO_2$, which are statistical results in one dimension. Because of the porous nature of the kerogen matrix, its density profile exhibited a high degree of heterogeneity. The kerogen matrix and $CO_2$ exhibited opposite distributions, indicating that the $CO_2$ component tends to reside in the low-density region of the matrix. The density profiles of the $CO_2$ component were predicted using the MDMC algorithm, and their results are in good agreement with the MD results. Fig. 6 illustrates the two-dimensional density contours of the MD and MDMC methods in addition to their one-dimensional distributions. Evidently, the MDMC method provides an accurate representation of the spatial distribution characteristics of the MD results. The high-density region shown in Fig. 6 represents the enrichment of $CO_2$ molecules, which correlates with the pore space in the kerogen matrix. In most cases, the low-density region corresponds to the skeleton of the kerogen matrix as $CO_2$ molecules cannot access every pore space.

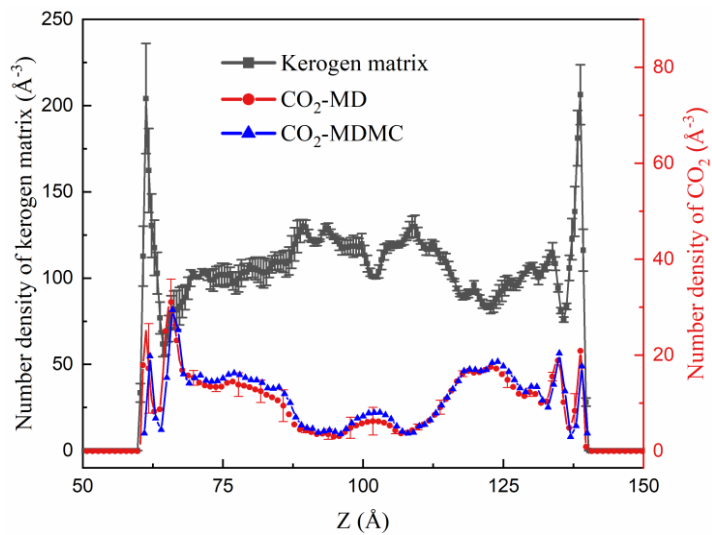

Fig. 5 One-dimensional spatial distributions of the kerogen matrix and $CO_2$.

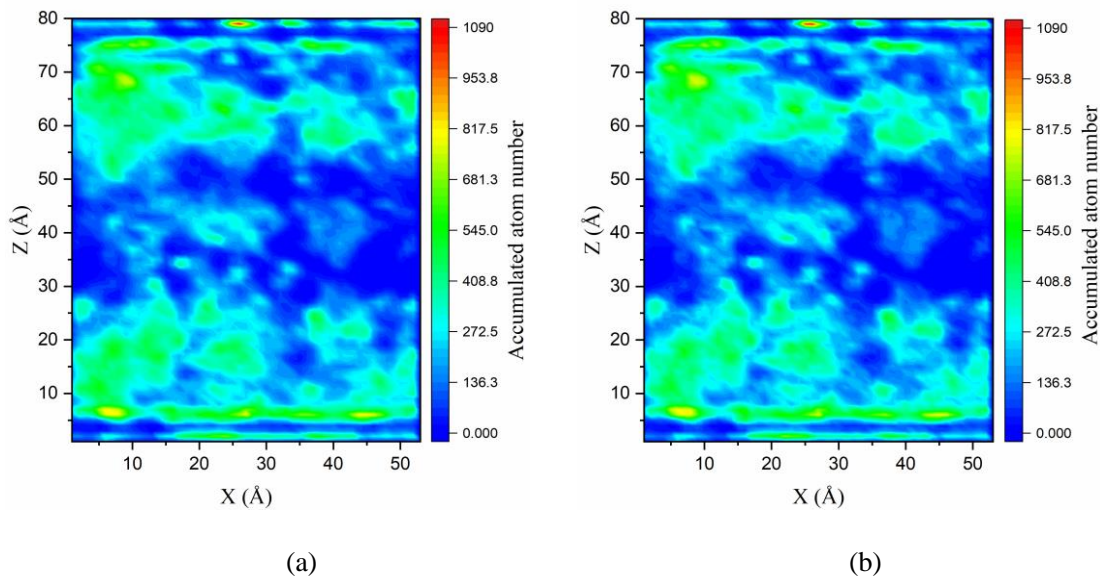

|       (a)       |       (b)       |

Fig. 6 Two-dimensional spatial density contours of $CO_2$ obtained from (a) the MD and (b) MDMC methods.

In the one-dimensional and two-dimensional simulation results, the MDMC method could provide good descriptions. Additionally, three-dimensional spatial distribution was able to offer more information about the $CO_2$ distribution and porous structure of the kerogen matrix. Each point in Fig. 7 represents a state that can be reached by $CO_2$ molecules in the MD and MDMC calculations. Fig. 8 illustrates the state agreement between the MD and MDMC methods, demonstrating that the MDMC algorithm has accurately captured the result from the MD simulation. The porous structure of the kerogen matrix is also visible in Fig. 7, demonstrating the heterogeneity of the pore space of the kerogen matrix, which induces an inhomogeneous $CO_2$ absorption.

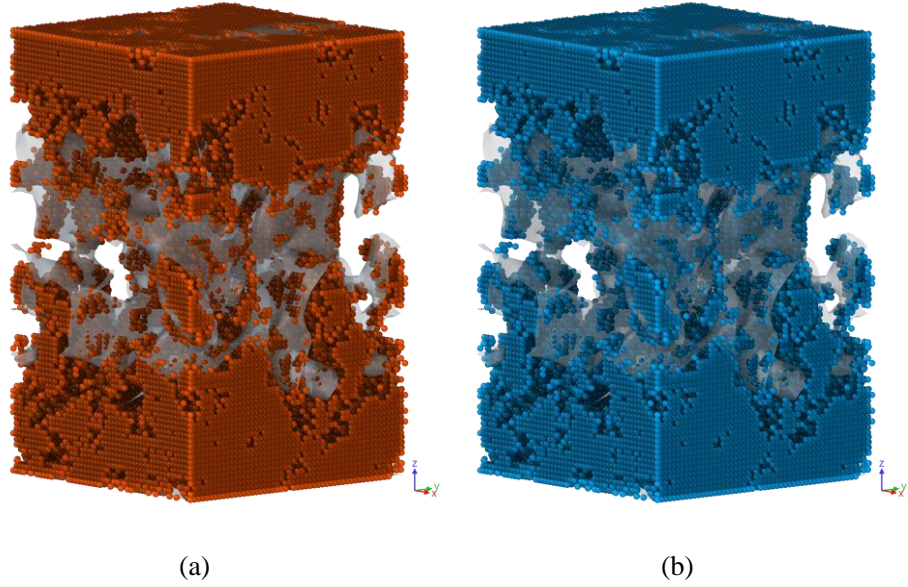

(a)                          (b)

Fig. 7 Three-dimensional spatial visualization of $CO_2$'s states using the (a) MD and (b) MDMC methods. Red particles represent MD spatial states, whereas blue particles show the spatial states predicted using the MDMC method.

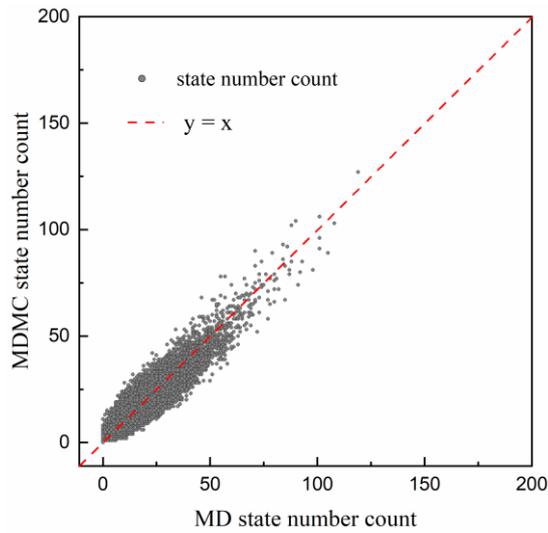

Fig. 8 Comparison of MD and MDMC results.

The coarsening operation can be presented in the MDMC method at both the spatial and temporal scales. In Fig. 9, the occupied volume of the $CO_2$ phase was used to evaluate the error of different coarsening operations. In order to make a more accurate comparison of the error evaluation, the mesh size and timestep in the MD result were set to one. Through the coarsening operations, the computation speed always increased, but the larger mesh size and timestep induced more errors. Consequently, an appropriate coarsening operation could be employed as long as the simulation error is acceptable. Despite this, the MDMC algorithm without coarsening was still significantly faster than the classical MD simulation, as shown in Table 5, and the MDMC algorithm was far more scalable for various problems.

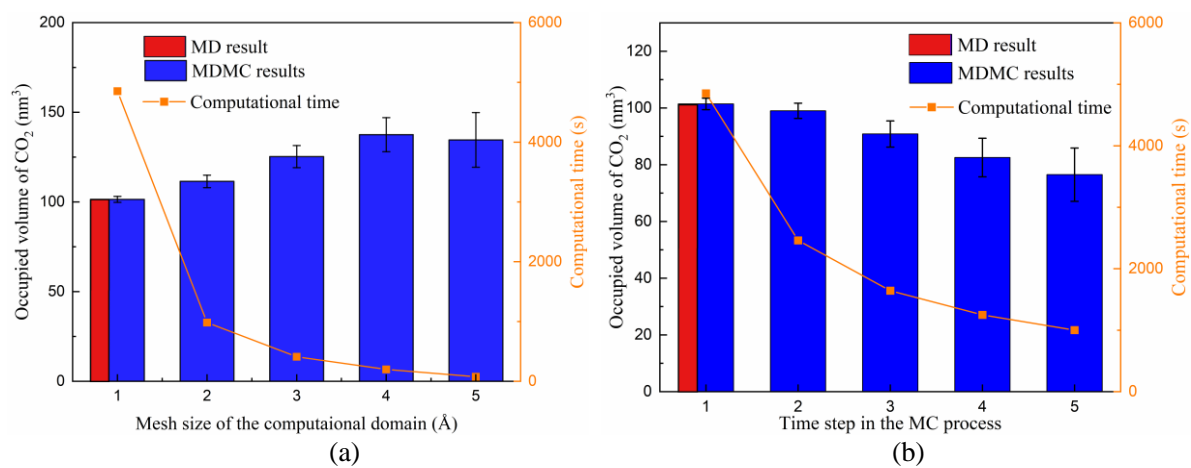

Fig. 9 Coarsening evaluation of the MDMC method in the (a) spatial and (b) temporal scales. All cases were calculated using Matlab codes on Intel Xeon CPU E5-2697 v2.

Table 5 Computational performance of the MD and MDMC methods.

| State of compression | MPI task | Computational time [s] | Speed up |
| --- | --- | --- | --- |
| MD | 96 | 60755 | 1 |
| MDMC (mesh size = 1.0) | 1 | 4852 | 1200 |
| MDMC (mesh size = 2.0) | 1 | 982 | 5942 |
| MDMC (mesh size = 3.0) | 1 | 410 | 14227 |
| MDMC (mesh size = 4.0) | 1 | 199 | 29309 |
| MDMC (mesh size = 5.0) | 1 | 76 | 76742 |
| MDMC (timestep = 1.0) | 1 | 4852 | 1200 |
| MDMC (timestep = 2.0) | 1 | 2460 | 2371 |
| MDMC (timestep = 3.0) | 1 | 1642 | 3552 |
| MDMC (timestep = 4.0) | 1 | 1249 | 4665 |
| MDMC (timestep = 5.0) | 1 | 1001 | 5827 |

*MD simulations were calculated using LAMMPS codes on Inter(R) Xeon(R) E5-2698 v3; MDMC simulations were calculated using Matlab codes on Intel(R) Xeon(R) E5-2697 v2.

### 3.2 Porous structural analysis of the kerogen matrix

As discussed in Section 3.1, the absorption behavior of the $CO_2$ component exhibited strong heterogeneity because of the porous structure of the kerogen matrix. This necessitated the study of the variation of the kerogen matrix.

Because of physical compaction, the buried depth of shale reservoirs significantly impacts the rock porosity. With increasing shale burial depth, rocks become denser and the preservation conditions for shale energy become better. Moreover, this increasing depth provides good conditions for $CO_2$ sequestration. Using the operations in Table 3, different compression states of the kerogen matrix were simulated. Fig. 10 shows the changes in pressure and temperature during the compression of the matrix along the entire simulation process. As a result of the complex molecular structure of kerogen monomers, compression will induce strong force within a short period, subjecting the pressure curves of the kerogen matrix to violent fluctuations during the simulation process. In most instances, shale reservoirs are located at depths exceeding a few thousand meters, resulting in very tight formations that maintain pressure around states 5 and 6.

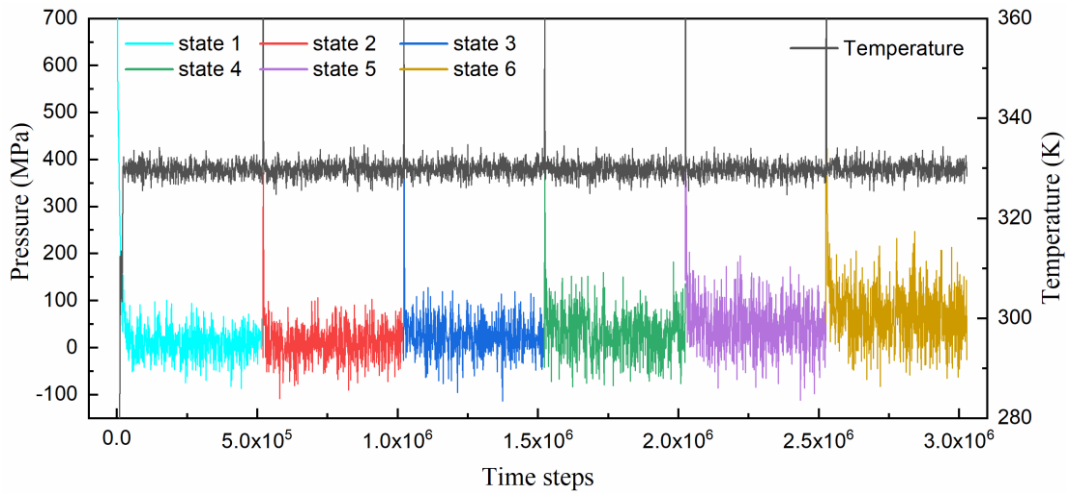

Fig. 10 States of pressure and temperature during compression processes. (T = 330 K)

Fig. 11 shows the PSD of the kerogen matrix for six states. As the compression process continued, the number of large pores, ranging from 23 to 13 Å, dwindled; whereas the PSD was mainly concentrated in a range of 5–10 Å. Beyond that, PSDs changed their characteristics from being dominated by large pores to three parts of pore size (0–5, 5–10, and 10–15 Å). In state 6, for instance, large pores (10–13 Å) were obtained through the contribution of dummy particles and medium and small pores were generated by the kerogen's complex molecular structure. $CO_2$ molecules were feasibly adsorbed in larger and medium pores but not in small pores as the latter were extremely difficult to access and cannot be used as effective absorption spaces unless the molecules were thermally moving over a long period of simulation time. The various PSDs quantitatively described the heterogeneous kerogen porous structure under different shale reservoir conditions.

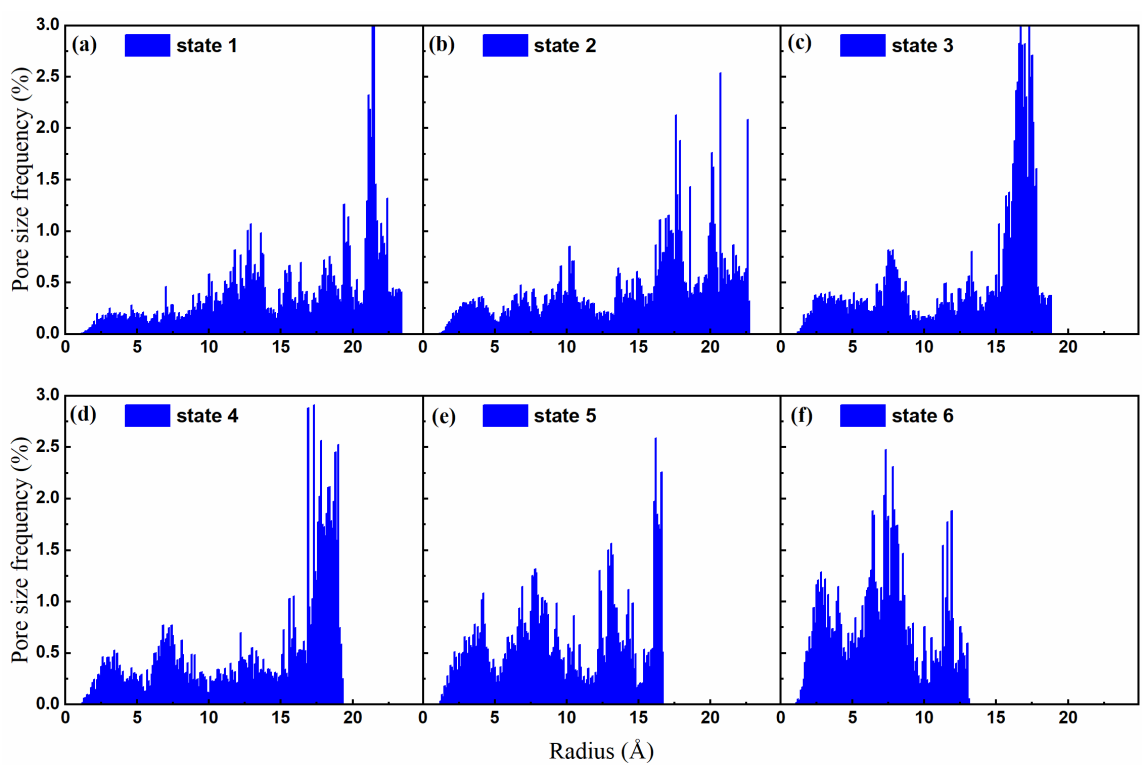

Fig. 11 PSDs during compression processes.

The surface area and accessible volume also influenced the $CO_2$ absorption. As shown in Fig. 12, the surface area and accessible volume of the kerogen matrix decreased, but the curve of the surface area showed a more lagged decline than that of the accessible volume. The reason was that most of the free volume was compressed in the initial state and not too much surface was in contact with each other, which resulted in a slow reduction of the surface area curve. Thereafter, as discussed in Fig. 11, the large reduction in the number of pores caused a sharp decline in the surface area. Moreover, strong compression state is not conducive to storing $CO_2$ as it leads to reduced pore space. Meanwhile, the compressed pores enhanced the interaction between $CO_2$ and the pore surface, which facilitated the absorption of $CO_2$. The density of the kerogen matrix also inevitably increased during compression. The density values of states 5 and 6 in Fig. 12 were consistent with those in previous studies [18-20], indicating that they represented proper density conditions for the kerogen matrix.

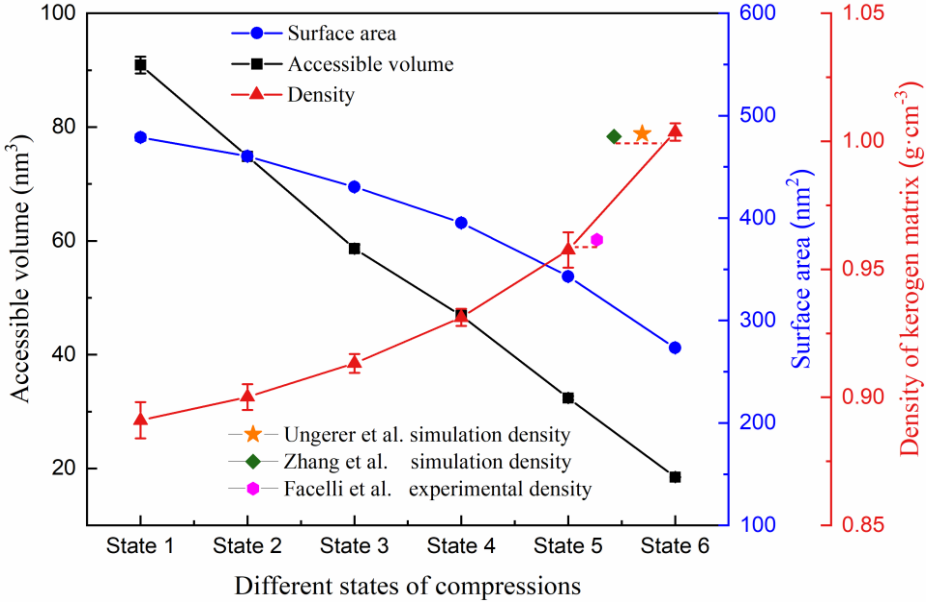

Fig. 12 Accessible volume, surface area, and density of the kerogen matrix during compression processes.

Interaction force contours were also addressed for different states, as shown in Fig. 13, to verify the rationality of the results concerning the energy mechanism. Overall, the interaction force increased from states 1 to 6. Because of the compression effect on the pores, local areas of high interaction forces were observed in states 2 and 3. When it comes to state 6, the interaction force was more homogeneous and higher. As a consequence, $CO_2$ absorption might be inhibited under high compression conditions according to the results of the interaction force.

Nevertheless, as discussed in Section 3.1, the kerogen matrix presented a porous structure. Thus, potential energy contours were computed to validate the precondition for $CO_2$ absorption, as shown in Fig. 14. In contrast to the contour of the interaction force, the potential energy distribution exhibited a strong heterogeneity. Because of the skeleton of the kerogen matrix and its self-assembly characteristics, the area of high-potential energy was presented as a banded distribution [18], which promoted the formation of the trapping region with low-potential energy. Therefore, although the strong compression resulted in high interaction force and tight kerogen matrix, the porous space with low-potential energy remained, allowing $CO_2$ molecules to be absorbed.

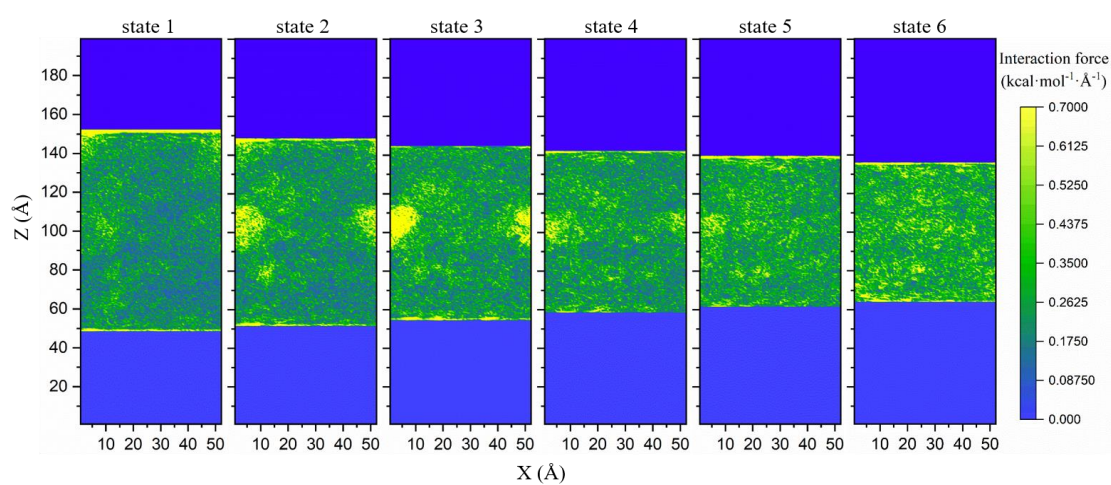

Fig. 13 Interaction force contours of the kerogen matrix during compression

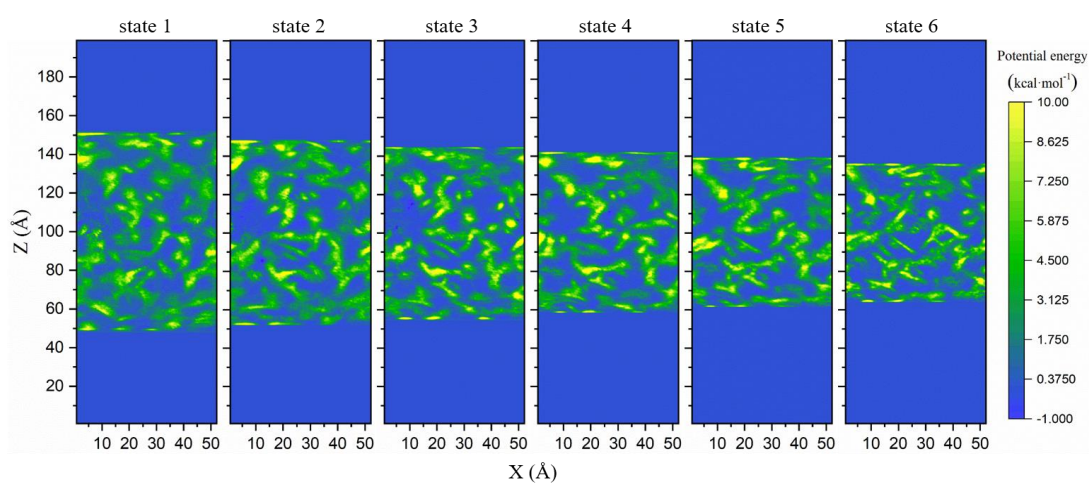

Fig. 14 Potential energy contours of the kerogen matrix during compression processes.

## 3.3 $CO_2$ absorption and its effect on the kerogen matrix

$CO_2$ absorption, as well as its effects on the kerogen porous medium, can be studied once the target state of the kerogen matrix is reached. According to Fig. 15, when $CO_2$ molecules were pushed into the kerogen matrix, the stress increased rapidly. A relaxation process was then presented to maintain the system at a lower energy level, and results were collected following the relaxation process. Consequently, the porous structure of the kerogen matrix was affected by the higher stress caused by the absorption of $CO_2$. This effect on the kerogen matrix could not be observed if the kerogen matrix was kept as a rigid body.

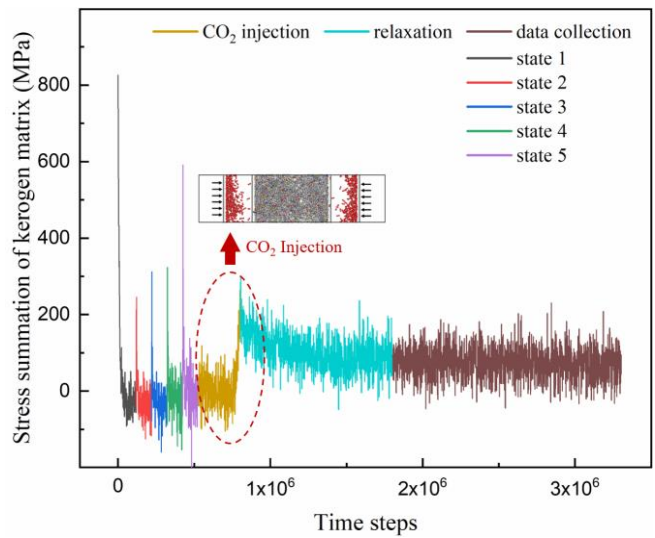

Fig. 15 Stress variations of the kerogen matrix in the process of $CO_2$ injection and absorption.

The potential energy surface of the kerogen matrix was determined both in the $X$ and $Y$ directions to validate the feasibility of $CO_2$ entry into the kerogen porous structures. Fig. 16 illustrates the heterogeneous distribution of potential energy, where high-energy areas prevented the entry of $CO_2$, whereas low-energy portions provided the pathway of access to the gas. Therefore, Fig. 16 indicates that the kerogen matrix always contained nanopores that facilitated $CO_2$ absorption, even though it presented as a very tight porous medium overall. This is consistent with the conclusion stated in Section 3.2.

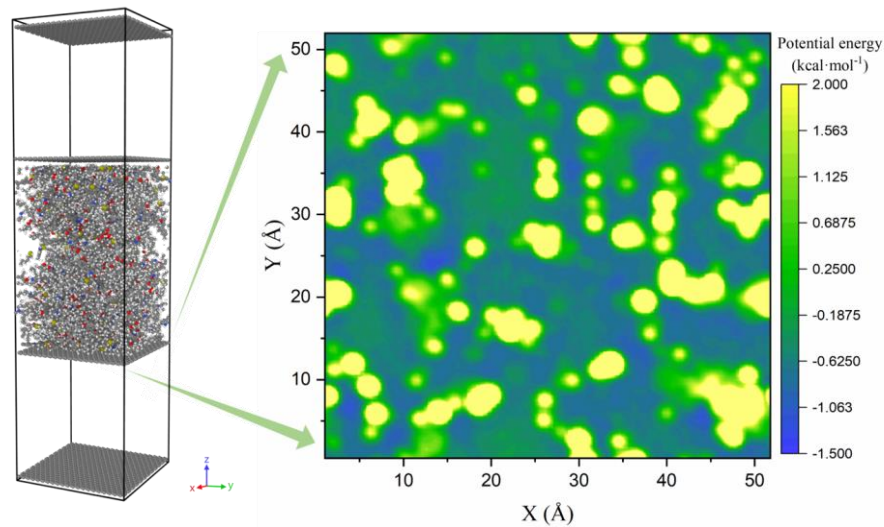

Fig. 16 Potential energy surface of $CO_2$ on the surface of the kerogen matrix.

The properties of kerogen porous media presented in Fig. 17 are used for investigating the effect of $CO_2$ absorption on the kerogen matrix. The PSD distribution in Fig. 17 (a) demonstrates a decline in small pores (0–3.5 Å) after $CO_2$ absorption due to the absorption of $CO_2$ into these small pores because of its thermal motion. In addition, because of space constraints, the small pores had a strong attraction for $CO_2$ molecules than the larger ones [9]. Consequently, the $CO_2$ molecules preferred to concentrate in smaller pores, causing them to swell and subsequently alter the skeleton structure. Following this, the large pore spaces were further compressed, resulting in a decline in the number of large pores. This process is clearly illustrated in Fig. 17 (c). Additionally, the variances of the PSDs before and after the $CO_2$ absorption are calculated to describe the degree of variation. The variance was reduced after $CO_2$ absorption, suggesting that the PSD was more average when the porous media absorbed$CO_2$ molecules. Because of this more averaged PSD, some numerical methods could reduce the computational error associated with $CO_2$ storage capacity. Fig. 17 (b) shows the changes in volume and porosity following the $CO_2$ absorption. The occupied volume of the kerogen matrix decreased and the accessible volume increased because of $CO_2$ absorption. This showed that the kerogen skeleton was compressed, resulting in higher porosity, which also corresponds to the above discussion.

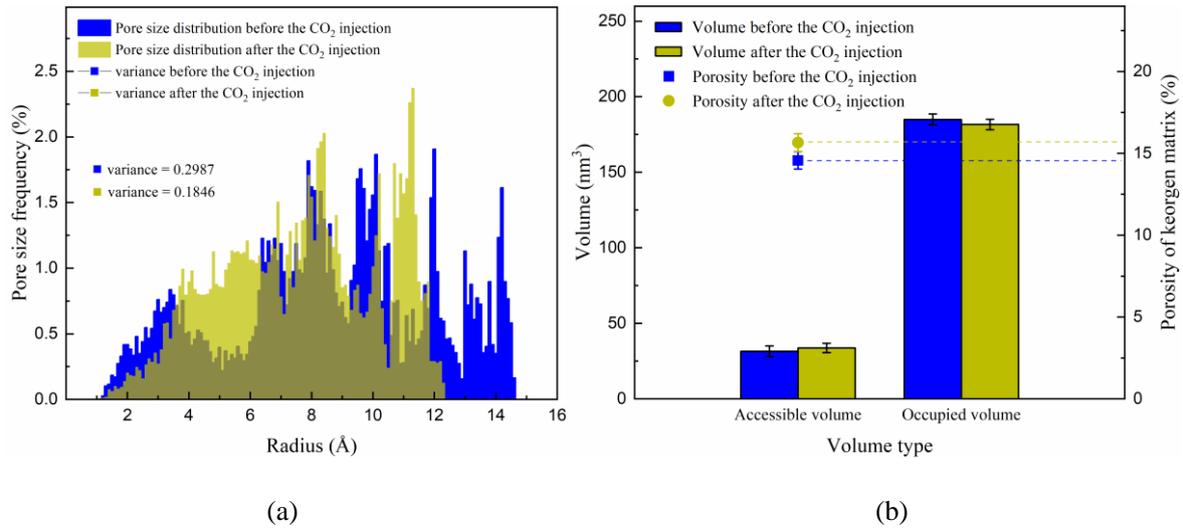

(a)　　　　　　　　　　　　　　　　　　(b)

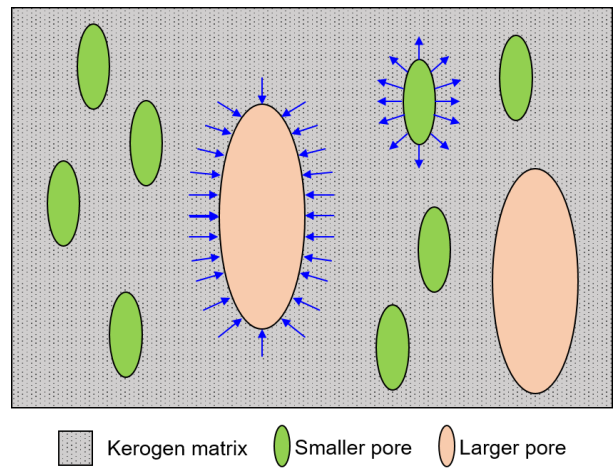

(c)

Fig. 17 (a) PSDs and (b) accessible volume, occupied volume, and porosity of the kerogen matrix before and after $CO_2$ injections. (c) Diagram of the PSD changing mechanism.

**3.4 Analysis of free energy at various temperatures**

The free energy of $CO_2$ molecules determines the thermal motion at which they move in the kerogen matrix, and the motion of $CO_2$ molecules is crucial to absorption processes. Consequently, the potential of mean force (PMF) was determined at different temperatures to investigate the absorption behavior of $CO_2$ in the aspect of energy mechanisms. The umbrella sampling method was used to calculate the PMF of $CO_2$, and the weighted histogram analysis method was applied to analyze the results [47, 48]. The simulation contained 40 windows with a range of 0.1 nm, and each window took 0.2 ns to reach equilibrium. The target $CO_2$ molecule was constrained by a spring force until it reached the kerogen matrix. The target $CO_2$ molecule was controlled only in the *Z* direction to obtain highly stable result [26].

Fig. 18 shows the different PMF curves of $CO_2$ molecules as they approached the kerogen matrix, and a yellow and gray transparent color scheme was used to depict two types of regions ($CO_2$ and kerogen matrix), respectively. Herein, the PMF curves were able to be classified into two parts: the $CO_2$ and kerogen phases. The temperature change did not appear to impact energy states in the $CO_2$ phase, but the positions of PMF with the lowest energy states were closer to the kerogen matrix at high temperatures. At the part of the kerogen matrix, high temperature facilitated the reduction of the $CO_2$ molecules' absorption resistance and resulted in a low free energy state. This suggests a highly stable absorption state for $CO_2$. Fig. 19 shows that high temperature plays a positive role in the volume fraction of $CO_2$, where $CO_{2\text{volume-fraction}} = CO_{2\text{volume-occupied}}/CO_{2\text{volume-accessible}}$, suggesting that the strong thermal motion of the kerogen matrix facilitated $CO_2$ molecules to reach more pores. The porosity presented a gentle decline due to the thermal expansion of the kerogen matrix. Therefore, the results in Fig. 19 have an excellent agreement with the PMF analysis in Fig. 18, and we can conclude that $CO_2$ storage has a better performance in reservoir with higher thermal sources. However, the development of geothermal energy could make the relevant area not suitable for $CO_2$ sequestration.

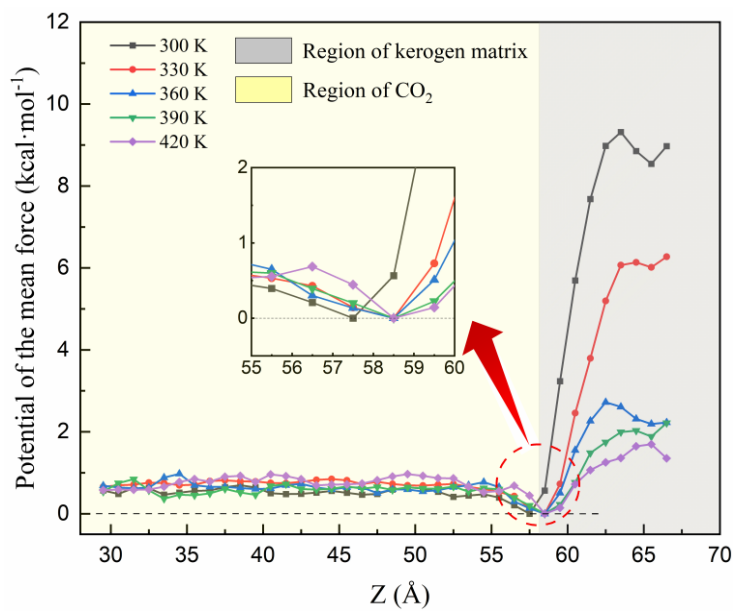

Fig. 18 Potential of the mean force between the $CO_2$ and kerogen matrix at different temperatures.

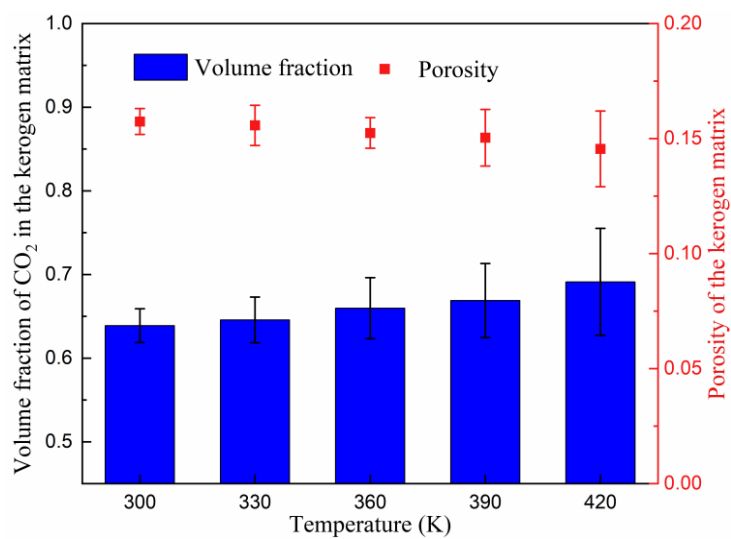

Fig. 19 Volume fractions of $CO_2$ in the kerogen matrix and porosities of the kerogen matrix under different temperatures.

## 4 Conclusions

In this work, the $CO_2$ absorption behavior in kerogen nanoporous media was studied using MD and MDMC methods. The results produced from MD simulations were taken as the ground truth for predicting the spatial distributions of $CO_2$ and matrix using MDMC algorithm. In terms of spatial distribution, the MD and MDMC methods showed good agreement in different dimensions, and the coarsening of space and time was examined in terms of computational speed and error. Although the coarsening operation induced computational errors, a much faster computational speed could be achieved even with a small mesh size and timestep compared with those of the classical MD method.

The structural properties of the kerogen matrix were investigated at different compression conditions, followed by analyses of PSD, pore volume, and surface area of the kerogen porous media. The PSD and density of kerogen matrix in highly compressed states agreed more with previous experimental and simulative studies. The interaction contour showed a highly compressed state, while the high-potential kerogen skeletons also constructed low-potential regions that provided accessible diffusion routes for $CO_2$. According to the interaction contours, the $CO_2$ entry was impeded by the highly compressed state, which was a negative factor, whereas the high-potential kerogen skeletons created low-potential regions that provided access to diffusion routes for $CO_2$.

Heterogeneous PES on the $CO_2$ entry surface revealed the accessibility of $CO_2$ in the kerogen matrix. This corroborates the above conclusions. Small pore spaces were enlarged by absorbed $CO_2$ molecules as $CO_2$ molecules were strongly attracted to small pores, whereas the large pores were consequently compressed. Furthermore, the kerogen matrix was compressed by the absorbed $CO_2$ molecules, resulting in an increase in its porosity that facilitated the sequestration of $CO_2$. Hence, the PSD was dynamically justified by the absorbed $CO_2$ and a highly concentrated PSD was obtained after $CO_2$ absorption, providing a significantly reliable reference for evaluating $CO_2$ storage ability. Because $CO_2$ absorption and diffusion were driven by thermal motion, the PMFs of $CO_2$ were calculated at different temperatures. At high temperature, $CO_2$ is able to overcome the kerogen's energy barrier and attained a highly stable energy state during absorption. Therefore, in scenarios such as the development of geothermal energy, the relevant area may not be suitable for the sequestration of $CO_2$.

**Acknowledgments**

We would like to express our appreciation for the following financial support: National Natural Scientific Foundation of China (Grants No. 51936001) and King Abdullah University of Science and Technology (KAUST) through grants BAS/1/1351-01 and URF/1/5028-01. For computer time, this research used the resources of the Supercomputing Laboratory at King Abdullah University of Science & Technology (KAUST) in Thuwal, Saudi Arabia.